# Chapter X
# Self-organization in embryonic development: myth and reality

Stuart A. Newman

**Abstract** "Self-organization" has become a watchword in developmental biology, characterizing observations in which embryonic or induced stem cells derived from animals replicate morphological steps and outcomes seen in intact embryos. While the term was introduced in the 18$^{th}$ century by the philosopher Immanuel Kant to describe the goal-directed properties of living systems, it came into modern use for non-living materials in which complex forms and patterns emerge through dynamical, energy-expending physical processes. What is the relationship among these uses of the term? While multicellular forms arose dozens of times from single-celled organisms, only some of these undergo development, and not all developmental processes are self-organizing. The evolution of the animals (metazoans) from unicellular holozoans was accompanied by the addition of novel gene products which mediated the constitution the resulting cell clusters as liquid-, liquid crystal-, and solid-like materials with versatile morphogenetic propensities. Such materials variously exhibited multilayering, lumen formation and elongation, echoing the self-organizing properties of nonliving matter, "generic" based on such parallels, though with biologically based subunit properties and modes of interaction. These effects provided evolutionary starting points and templates of embryonic forms and morphological motifs of diverse metazoan lineages. Embryos and organ primordia of present-day animal species continue to generate forms that resemble the outcomes of these physical effects. Their development, however, employs overdetermined, highly evolved mechanisms that are often disconnected from their originating processes. Using the examples of gastrulation,

Stuart Newman
New York Medical College
Valhalla, New York USA
Stuart_newman@nymc.edu

somitogenesis, and limb skeletal development, this chapter provides instances of, and a conceptual framework for understanding, the relationships between transparently physical and evolved types of developmental self-organization.

## X.1 Introduction

A series of findings beginning in 1980s, including ones arising from improved cell culture methods, comparative phylogenomics, cellular and genetic manipulation of embryos, and progress in understanding the dynamical and other physical properties of multicellular materials, led to the mostly unacknowledged disappearance of the notion of a "genetic program" from the theoretical discourse in developmental biology. This idea, inspired by the rise of the digital computer in the 1950s and the associated hardware-software distinction, attempted to locate the information acquired during phylogeny in each organism's nuclear DNA, where it was deployed in a hierarchical fashion during ontogeny (Istrail, De-Leon, and Davidson 2007, Peluffo 2015; Sarkar 1998). What increasingly replaced the computer program model was the concept of "self-organization," whereby complexity of form and pattern (usually, but also potentially of function) emerges from apparent simplicity due to interactions of unorganized components. This term had already been in use in the physics of materials of the meso-, or middle-scale, "soft matter" (de Gennes 1992) such as viscoelastic fluids, and "excitable media" such as complex chemical and mechanochemical systems (Elphick et al 1990). In a sense, the conceptual change echoed the centuries-old transition from preformationism to epigeneticism that had been played out in the pre-genetic, pre-evolution era of biology (Van Speybroeck et al 2002).

Some findings from the fields mentioned above which led to changes in how development and its evolution were conceptualized included:

(i) The capacity to generate and maintain stably differentiated cell types, as well as early embryo- and organ-like structures, in culture. These depended on identification of soluble growth factors such as Wnt, TGF-βs and FGFs and extracellular matrix (ECM) molecules such as collagens, fibronectin, and laminin (Gospodarowicz 1984; Gospodarowicz et al 1986; Hay 1991 ; Massagué 1987). Later it was shown that members of the same set of factors could elicit differentiated cell types and early embryonic structures ("gastruloids") from both embryonic stem (ES) cells and induced pluripotent stem (iPS) cells (Beccari et al 2018; Hanna et al 2008; Hoffman and Carpenter 2005).

(ii) Advances in sequencing techniques and comparative phylogenomics, which led to the recognition that these same growth and ECM molecules were among a few dozen conserved members of a phylogenetically conserved "developmental genetic toolkit." The products of these genes were found to induce changes in form and pattern in embryos across the full range of animal phyla (Carroll et al 2004).

(iii) Visualization of gene expression in whole embryos, which showed that given toolkit gene products could be involved in the formation of morphological characters that were both homologous (e.g., the legs of different vertebrates) and analogous (e.g., the legs of mice and of flies) across phyla. This "homology-analogy paradox" (i.e., anatomical analogs are not homologs, but the molecules that mediated both could be homologous) was difficult to reconcile with the widely accepted notion of the evolutionary continuity of purported genetic programs (Newman 2006).

(iv) Mathematical and computational models of cell movement and differentiation in response to toolkit gene products, which consistently found that physical determinants such as differential adhesion, reaction-diffusion ("Turing-type") processes, and oscillations in gene expression provided the most coherent and convincing explanations of the relationships between gene action and morphological phenotype (Forgacs and Newman 2005).

(v) Genetic manipulation, including full and conditional knockouts and overexpression, indicated that gene function could be decoupled from morphological outcome in ways difficult to reconcile with the program notion. The discovery of "constructional unit autonomy" and "developmental system drift" pointed to a similar phenomenon of gene-morphology decoupling over evolution (Müller and Newman 1999; True and Haag 2001;).

(vi) The formation of chimeras from mammalian species as evolutionary distant as sheep and goats, which diverged 14-16 Mya by some estimates. Blastomeres of these species could be mixed together to produce healthy full-term animals of intermediate phenotype (Fehilly et al 1984). This was exceedingly difficult to reconcile with a genetic program notion since chimeras are cellular mosaics of the two originating species. The cells (unlike those of hybrid organisms) retain their species-specific genomes along with whatever additional cytoplasmic information is used for development and read each other's signals in constructing an evolutionarily unprecedented animal. No conventional definition of computational programs would admit of this kind cooperation among fortuitously associated entities.

Despite this evidence against the genetic program notion, the concept of "self-organization" as applied to embryogenesis or organogenesis is nonetheless ambiguous. It is not identical to the claim that physical processes underlie morphogenesis, which is an uncontested assumption of modern biology. Something more is implied by using the term "self-organization" instead of simply "development." Machines are physical entities contrived to produce specific outcomes. While organisms and their constituent processes are not machine-like (Nicholson 2019), their structure-function relationships are often difficult to discern, not least because their protein constituents are often intrinsically disordered and alter their roles based on conditional interactions with other components (Uversky and Giuliani 2021; Niklas et al 2015). Machine analogies are often a recourse of convenience.

Complex materials can undergo morphological changes over time and arrive at a reproducible endpoint that might not resemble anything else in nature. This is true of certain minerals, as well as the developing tissues of plants and fungi. The morphogenetic mechanisms that produce flowers, pine cones, and gilled mushrooms are poorly understood, but there is no reason to doubt that are physical processes. Under certain conditions they produce patterns of exquisite mathematical regularity that have plausible physical interpretations (Douady and Couder 1992; Azpeitia et al 2021). It is possible that these systems are "self-organizing" in the sense of being mesoscale, excitable materials with physically explicable, reliable outcomes. But the enigmatic physical nature of plant and fungal tissues as growing, deformable solids makes this difficult to establish.

Physical organizing processes and effects, and corresponding morphological outcomes, have been termed "generic" if they are common to both living and nonliving materials (Newman and Comper 1990). While even non-generic physical processes can be self-organizing, it is only for systems with generic properties that self-organization can feasibly be confirmed to be an important mode of development. This is because generic processes are governed by known physical principles with predictable outcomes. The possibility that generic processes were active during the periods when a lineage's morphology first arose in evolution can also provide insight into why they look the way they do.

In this chapter I define "physical self-organization" (or "self-organization" for short) in conformity with the classical usage in physics: the generation of geometric structures or chemical patterns by dynamical interactions of unorganized components. The outcomes of such processes, also referred to as "dissipative structures" (Goldbeter 2018), only occur in thermodynamically open systems, in which there are fluxes of matter or energy across the boundaries. Physical self-organization is distinguished from "self-assembly," the fitting together of a system's components (as in protein quaternary structures) that does not involve such system-environment exchange (Halley and Winkler 2008). Since, as noted, generic processes pertain to living as well as nonliving systems, in the former cases, where the generic physical mechanism (e.g., the random motion of subunits in a liquid-like material; see below) depends on a biological, rather than a nonbiological, function (undirected cell motility vs. Brownian motion in this example), the term "biogeneric" is used (Newman 2016a).

In contrast to multicellular plants and fungi, aggregative microorganisms such as the cellular slime mold *Dictyostelium discoideum* and the bacterium *Myxobacteria xanthus* present uncontroversial examples of physical self-organization, employing generic processes (Fujimori et al 2019; Fukujin et al 2016; Wu et al 2011). But even in these seemingly simple systems, interpretation of morphogenetic processes is confounded by the capacity of individual cells to act as autonomous agents, with apparent goals of their own, a feature not seen in nonliving matter (Arias Del Angel et al 2020).

Regarding the animals, the subject of this chapter, the liquid-like nature of their embryonic tissues imposes coherency and collective behaviors on their cells. This affords possibilities of generic-type self-organization, as does the ability of such tissues to serve as excitable chemical and electrical media. In the discussion here, processes internal to individual cells, including those driving their random motion or regulating the expression of genes, and the bases of fluidity and excitability, will be "bracketed" with no attempt to explain them. Development will only count as self-organized in this framing if it is attributable to physical processes that act on the multicellular, rather than the subcellular level.

Newer in vitro culture techniques which permit isolated embryos or aggregates of embryonic or induced pluripotent stem cells to develop along normal pathways have permitted detailed investigations of developmental capabilities and subsystem interactions. Development of "embryoids" or "gastruloids" initiated from stem cells under these conditions is often taken as evidence of self-organization (Bedzhov and Zernicka-Goetz 2014; Beccari et al 2018; Etoc et al 2016). But simply because an embryo can be removed from its normal gestational setting and in the presence of appropriate nutrients execute the same set of steps it does in vivo, does not mean it is exhibiting physical self-organization. It could be behaving as a more complicated (because highly evolved) or causally obscure (see below) physical assemblage.

This is true even if the starting point of in vitro embryogenesis is a collection of embryonic ES or iPS cells. After all, such cells are derived from organisms that emerged as distinct mammalian species millions of years ago and retain their respective species-specific genomes when reverted to the stem-state. It would be surprising if they had not evolved the means to recover the morphogenetic routines characteristic of their type. The kind of self-organization exhibited, however, is not the physical variety, for which complexity is a truly emergent effect. It is rather the restorative behavior of a highly evolved, complex "self."

While the fact of evolution thus complicates the ascertainment of self-organization, the possibility of such effects, particularly if they have generic properties, can provide insights into the processes of evolutionary innovation. In the older "genetic program" picture, an organism's developmental mechanisms were presumed to have evolved gradually by random mutation and selection, with no role for the material properties of the forming tissues. Although organisms with programmed development could potentially result from such a random search, they might resemble the self-constructing contrivances theorized by the mathematician John von Neumann (Von Neumann and Burks 1966). Such machines would be vanishingly likely to arise and could only operate to produce the requisite forms when all the parts snapped into place. In contrast, if self-organization of tissue masses play a part in embryogenesis, or once did in a lineage's phylogenetic history, pathways of morphological evolution would have had preferred directions and outcomes, contrary to the logic of undirected Darwinian natural selection.

In the next section, I show how physical self-organization in fact helped to kick-start the evolution of developmental systems. I describe how animals and their tissues arose from ancestral holozoan cells that were recruited into multicellular entities by the products of a specific, evolutionarily assembled genetic toolkit. These masses of cells constituted parcels of novel fluid-like living matter with unique morphogenetic propensities. In the subsequent section, I discuss three examples from vertebrate embryogenesis: gastrulation, somitogenesis, and limb skeletogenesis, which have been proposed, with experimental evidence, to employ various generic physical processes to implement their morphological outcomes: differential adhesion, a biochemical oscillation in conjunction with a molecular gradient, and a Turing-type reaction-diffusion standing wave-generating system.

My goal is to show that whereas generic physical processes typically continue to be involved in the developmental events, they are not always, or equally in all relevant species, the decisive factors in morphogenesis. In consequence, they are not generally explanatory of the observed developmental phenomena. If this is correct, descriptions of purported self-organization of stem cell embryoids and gastruloids, which are increasingly prominent in the developmental biology literature, are misleading regarding mechanistic understanding. In the concluding section I suggest that only with an analysis incorporating an evolutionary dimension, i.e., a recognition of how development in an extant species reflects the originating processes of its lineage followed by their reconfigurations and transformations, can we hope to comprehend in causal terms how the morphological phenotype is realized.

## X.2 Metazoan matter: physical bases of self-organization

For the concept of "self-organization" to be nontrivially different from "development," the forming structures should emerge de novo from unstructured precursor materials by known effects with generally predicable outcomes. The non-self-organizational alternative would be the playing out of the morphogenetic potential of a complex, composite material that has evolved to unfold in certain ways. Animal embryos generally fit the second description. Do they partly conform to the first?

The animals are members of Holozoa, a phyletic group which also includes unicellular and transiently colonial organisms. Premetazoan holozoans are inferred to have contained cadherin-like cell surface proteins (Nichols et al 2012). In the transition to the metazoan form of life some of these proteins acquired a cytoplasmic domain that mediates attachment to the cytoskeleton. The resulting "classical" cadherins are exclusive to the animals, and some of the cytoplasmic adaptor proteins of the linkage are also animal-specific (Abedin and King 2008).

Materials composed of subunits (e.g., atoms, simple molecules, polymers, or cells) that have strong affinity for one another while readily changing their neighbors are defined as liquids (Widom 1967). The homophilic adhesive function of classical cadherins, with their unique transmembrane-cytoskeletal connection that permits the cells of metazoan embryos and organ primordia to move relative to one another while remaining attached to their changing neighbors, causes the cell clusters to behave like liquids (Foty et al 1994, Steinberg 2007, Miller et al 2013). Unlike nonliving liquids, however, where nanoscale subunits change position by Brownian motion (which is independent of their internal structure), for liquid tissues, the active mechanochemical processes of the (macroscopic) cells' interiors are essential to drive their random movement.

In determining whether embryos and organs develop by self-organization rather than by some alternative, complex, mechanistically obscure set of physical processes, we can ask whether the morphological features that appear during ontogenesis are among the inherent modes of this liquid-tissue "ground state." For example, in nonbiological liquids, mixtures of subunits which are sufficiently different in composition (e.g., adhesivity) can undergo phase separation, forming immiscible layers (e.g., oil and water) (Flory 1973). If the subunits have nonuniformly distributed interactive properties ("binding polarity") lumens or interior spaces can form. Further, liquids whose subunits have asymmetric shapes can form liquid crystals, droplets of which become oblate or elongated rather than spherical, as in ordinary liquids (Yang et al 2005).

All these liquid phenomena – multilayering, cavity formation, reshaping of masses, and some additional behaviors of this state of matter – spreading on (or "wetting") solid substrata, and "jamming," a liquid-to-solid phase transformation in which the subunits move closer to one another (Bi et al 2011), halting their relative movement – are features of embryonic development, though not all of them in every phylum. As noted, the subunits of liquid-like tissues are motile cells, and the gene products responsible for the strength of their mutual adhesion, and their binding and shape polarity are well known. These gene products did not appear all at once during evolution, but sequentially, with the present-day descendants of earliest emerging animals, marine sponges, placozoans, cnidarians, and ctenophores having some, but not all of them. Organisms in these groups are correspondingly morphologically simpler than the ones such as arthropods, mollusks, and chordates (including humans) having fuller genetic toolkits (reviewed in Newman (2019b)).

With demonstrable liquid-like properties and structural motifs that resemble those seen in liquids, it is reasonable to expect that animal embryos and organ primordia would generate their forms by known physical means. However, as mentioned above, this supposition is often incorrect. For example, development can deviate from generic physics owing to the capacity of the cellular subunits to actively participate in morphogenetic processes. Rather than simply moving in random directions (the prerequisite for their aggregates behaving as liquid-like materials) they act as signaling and reactive centers (Forgacs and Newman 2005) and differentiate into specialized types (Newman 2020).

The capacity of cells to differentiate is a feature of animal development that previously had been thought to be based on generic physical effects. Specifically, the panoply of specialized cell types – a few in sponges and placozoans and more than 200 in mammals – was proposed to be the mathematically determined dynamical modes of behavior of the respective genomes. Genes regulate other genes in all forms of life and are thus organized into gene regulatory networks, or GRNs. These systems can be modeled as Boolean switching networks (BSNs) in which genes turn one another on or off at successive discrete time steps based on the states of their input genes at the previous time step, or as ordinary differential equations (ODEs), where the both the time progression and concentration of gene products are continuous variables. Formally, both kinds of networks exhibit multiple stationary modes of activity or dynamical "attractors," in which the concentrations of the system's components (e.g., gene products or on-off states) are unchanging over time, or cycle among a small subset of the possible states of the system. Since transitions are possible between such attractor states, these generic mathematical representations were proposed to reflect the properties of genomic organization, with the dynamical attractors providing the explanation of the existence of cell types (Kauffman 1993).

More recent work, however, has led to the recognition that the physical basis of cell differentiation in animals is not generic in the sense defined here. Specifically, the system of expression hubs in which functionally related genes are partitioned into insulated nuclear regions (topologically associated domains: TADs) by a combination of the mechanics of chromatin fibers and phase transformations of associated transcription factors and scaffolding proteins (Furlong and Levine 2018; Newman 2020) has no non-biological counterparts. Some of the proteins involved, and the distant gene regulatory sequences known as enhancers, which congregate in TADs in up to the thousands in differentiated cells, are specific to metazoans, not even being present in other holozoans. The lack of stoichiometric and mass action relationships among cell-type-specific genes and their molecular regulators, as well as the transience of the causal nexuses in which they are involved, make BSN or ODE representations applicable to these GRNs at most on in a local (lineage-adjacent) sense, but not on the level of the global "regulatory genome" (Peter and Davidson 2015). Moreover, it is highly implausible that a system of physiologically coordinated cell and tissue types could be produced as a purely mathematical consequence of the balance of effects leading to dynamical attractors (Newman 2020).

Differentiated cell types, even if generated by processes that by most definitions are not self-organizing, can be induced and spatially patterned by self-organizing processes acting on a multicellular scale. The resulting arrangements, in turn, can trigger additional self-organizing processes. Some of these effects resemble ones seen in nonbiological systems. A simple case is the appearance in some embryos or developing organs of a specialized group of cells that secrete a "morphogen," a molecule that causes other cells to change their states depending on concentration or duration of exposure. The transport of the morphogen away from its source can be by diffusion (Yu et al 2009), a generic process, or by something more complicated and cell dependent (Ben-Zvi et al 2011). A more complex case involves the coupling of molecular transport (however mediated) across spatial domains with the biochemical responsivity (excitability) of cells of those domains. This can produce nonuniform patterns such as standing waves (i.e., spots or stripes) of molecular concentration, followed by periodic distributions of an induced cell type (Kondo and Miura 2010).

Processes of this sort (often referred to as "Turing-type" owing to a paper in which the mathematician A.M. Turing proposed them as "the chemical basis of morphogenesis"; Turing 1952), a paradigmatic example of self-organizing pattern formation in nonliving systems (Castets et al 1990; Boissonade et al 1994; Ouyang and Swinney 1991), have been proposed, for example, to underlie patterning of digits in the vertebrate limb (Newman and Frisch 1979; Zhu et al 2010), pigment stripes and spots of animal skins (Kondo and Asai 1995), and hair follicles (Sick et al 2006). As we will see below, however, actual developmental patterning mechanisms, while they may have originated in the form of these generic ones, are typically evolutionary transformations of these mechanisms into more elaborate, non-generic ones.

Other seemingly generic physical effects, e.g., phase separation of tissue layers by differential adhesion, segment formation by the interaction of synchronized biochemical oscillations and tissue gradients ("clock-and-wavefront" mechanisms) have also been transformed over evolution, and are differently realized in different lineages, sometimes being supplanted by very different mechanisms (Newman 2019a; Haag and True 2018). A theoretical rationale for such divergences is that if a layered, or a segmented morphology becomes intrinsic to the organism's identity, conserving the forms will impose constraints on genetic changes, leaving the structural motif (layers, segments, or endoskeletal elements such as digits) in place as the developmental process evolves into something more complicated or entirely different (Müller and Newman 1999; Newman et al 2006).

Finally, as mentioned, the living cells of developing embryos and organs, though participating the reorganization of liquid-like materials, are not physically passive like the subunits of nonliving liquids. The reactive and agent-like natures of these entities will therefore complicate attribution of generic self-organization to morphogenesis (Arias Del Angel et al 2020). This agential behavior has many intersecting determinants and an obscure evolutionary history. As noted above, autonomous cell behavior is important in the development of organisms, both non-metazoan eukaryotes and prokaryotes, in which multicellularity is achieved by aggregation,. In more complex forms in which multicellularity is achieved zygotically, individual cell agency is more restricted. In plants, although extensive cell-cell communication is employed via plasmodesmata during development and growth, the presence of encapsulating walls curtails cell autonomy and thus agency. In animal embryos, directed cell movement and other autonomous behaviors are also less prominent than in aggregative forms, but always remain a possibility in the generation of animal form and pattern.

## X.3 Evidence for and against developmental self-organization

In this section, I discuss three embryonic processes in animals: gastrulation, segmentation, and endoskeletal patterning, for which there is good evidence from in vitro studies for self-organization of the generic type. I will also describe evidence from in vivo, or other in vitro studies that challenges these attributions. Apart from these contrasts, I make no attempt to provide a detailed account of these developmental events. Each of these embryological phenomena occurs in invertebrate as well as vertebrate organisms, but the examples I explore are all taken from the latter group.

*Gastrulation*

Gastrulation is the formation and arrangement of distinct tissue layers during the establishment of the body plan in early animal development. About half a dozen distinct kinds of cell behavior are employed, depending on the species (see Fig. X.1, top panel, for the zebrafish case), and each requires a differentiation step that designates a subpopulation of the original clump of cells derived from the zygote (variously, blastula, morula, inner cell mass) as distinctive. Often the defining character is an adhesive differential that causes the subpopulation to be more, or less, cohesive than the other cells of the mass (reviewed in Newman (2016b) and Forgacs and Newman (2005)). How the different cell populations are established appears to depend on maternal determinants stored in the egg, or external cues during oogenesis or gestation. Self-organization may also be involved in the early stages of cell allocation: some evidence points to one or more Turing-type reaction-diffusion mechanisms (Muller et al 2012).

Steinberg and his colleagues, referring to the physics of phase separation, attributed the layering of germ layers in gastrulating frog embryos to measured cohesivity differences of the tissues. The interfacial tensions between the layers, which were the relevant determinants for this putative thermodynamically driven effect, could not be directly measured, but were inferred indirectly by comparing the surface tensions (measured by a compression tensiometer) of rounded-up parcels of isolated tissues (Foty et al 1994). The values confirmed the predictions of the Differential Adhesion Hypothesis (DAH), which, under the assumption that the tissues were liquid-like (which is generally correct, see previous section), asserted that surface and, by inference, interfacial tensions, were due exclusively to quantitative or qualitative differences between the tissues' cell types in adhesive proteins, such as classical cadherins (Steinberg 2007).

Experiments on sorting out of mixtures embryonic cells from later-stage embryonic primordia (liver, limb bud, retina, etc.) was also consistent with the DAH. Here, the equilibrium configuration of the tissues resulting from demixed cells from given sources was always identical to the morphological results of direct confrontation of the intact tissue fragments. This was true with respect to which mass was engulfing and which engulfed, or (where full engulfment did not occur) in the direction of convexity of the interface (Duguay et al 2003).

These experiments confirmed the liquid nature of the tissues, and their phase separation-like behavior (a generic physical effect). They did not prove the attribution of the tissue configurations to differential adhesion per se, however. This aspect of the hypothesis appeared to be pinned down, however, when Steinberg and Takeichi genetically engineered mouse L cells, a cell type with no adhesive proteins of their own, to express different amounts of P-cadherin (Steinberg and Takeichi 1994). Here, sorting out occurred when the levels of expression were sufficiently different, and the resulting artificially produced tissue-like masses exhibited engulfment behaviors and interfacial contours exactly as predicted by the DAH.

Cells, however, are different from subunits of nonbiological liquids in that they have mechanically complex, responsive interiors. Interfacial tension, the driving force for cell sorting and tissue engulfment in liquid tissues was postulated by the DAH to result simply from cell surface-based adhesive differentials between the cells. But Brodland countered this by observing that that the compression tests used by the Steinberg group measured not only the surface tension between the aggregate and the surrounding medium, but (indirectly) also the effective viscosity and contractility of the cells' cytoplasm, which affect the cohesivity of the aggregates. This led him to propose an alternative, differential interfacial tension hypothesis (DITH) for cell sorting and tissue engulfment in vitro and in vivo (Brodland 2002).

While adhesive differentials alone can be the determining factor for cell sorting or phase separation of tissue fragments if internal cell states are equivalent, as in the genetically engineered L-cells in the Steinberg-Takeichi experiments, the more general DITH was ultimately recognized (even by advocates of the DAH) as more relevant to embryonic systems (Manning et al 2010). In vitro experiments on tissues of gastrulating embryos appeared to confirm the idea that the boundaries of immiscibility between germ layers and the relative spatial arrangements of the layers were the result of differential tissue surface tension (TST)-based self-organization. Thus, progenitor cells from the ectodermal and mesendodermal germ layers of zebrafish embryos sort out from mixtures *in vitro*, and fragments of these layers adopt configurations that correspond to TSTs predicted from direct measurement of cell adhesivities and cortical tensions by atomic force microscopy (Krieg et al 2008). Moreover, the configuration of the leading edge of the spreading epiblast during epiboly (envelopment of the yolk cell mass) in zebrafish was consistent with the physics of surface wetting, i.e., TST-driven rearrangement (Wallmeyer et al 2018).

But what holds in vitro does not always reflect the in vivo process, as Winklbauer and coworkers noted when they reported "[c]adherin-dependent differential cell adhesion in *Xenopus* causes cell sorting in vitro but not in the embryo" (Ninomiya et al 2012). Further, the group that had previously obtained in vitro evidence for TST-driven rearrangement of germ layers (Krieg et al 2008) used optical methods to directly measure differences in germ layer TST when the germ layers are first established. They found, surprisingly, that the magnitude of this difference was insufficient to drive progenitor cell sorting and tissue rearrangement under physiological conditions (Krens et al 2017). The disparity between the *in vitro* and *in vivo* results was attributed to differences between the fluids bathing the cells under the two conditions. The low osmolarity of the interstitial fluid between the intact embryo's epiblast cells caused a reduction the interfacial tension between the tissue layers compared to that in the tissue cultures. These results suggest that the self-organized germ layer phase separation seen in vitro was artifactual. If so, a different explanation is required for this phenomenon.

The study's authors provided evidence that the germ layer interface was formed by directed cell movement, possibly under the guidance of a gradient of interstitial fluid (Krens et al 2017). Though biologically plausible, this explanation, which appeals to the agent-like properties of the embryo's constituent cells challenges the postulate of physical self-organization. It implies, instead, that the cells have an evolved capacity to assume their proper positions in the embryo based on something akin to a predetermined program.

But even if directed cell migration is involved in germ layer formation in zebrafish, it is not necessarily the mechanism for this developmental process in other species: there is extensive phylogenetic variability of gastrulation-related molecules and cell behaviors even among the vertebrates (Schauer and Heisenberg 2021). Moreover, the liquid-like properties of animal tissues are deeply entrenched in metazoan origins and the observed germ layer boundaries conform to the predictions of the physics of wetting (Wallmeyer et al 2018). One possibility is that the inferred program of directed migration may have been templated by, and over time supplanted, the formation of germ layers by the generic physical mechanism of tissue phase separation in a vertebrate ancestor (Newman 2019a). What is clear, though, is that for extant animals any genuine physical self-organizing processes that may act during gastrulation are integrated into more complex multidetermined developmental routines of these organisms, notwithstanding claims to have observed "self-organization" of gastrulation in vitro (Beccari et al 2018, Etoc et al 2016, Rosado-Olivieri and Brivanlou 2021).

*Somitogenesis*

In vertebrates, the mesoderm directly to either side of the notochord in the early embryo becomes organized into parallel blocks of tissue called "somites." The first somite forms as a tight aggregate or condensation of cells at the anterior region of the trunk. Each new somite forms immediately posterior to the previous one, budding off from the more anterior portion of the unsegmented presomitic mesoderm (PSM) (see Fig. X 1, middle panel, for the chicken case). The somites mature into the vertebrae and the ribs, and the associated muscles. They also send muscle progenitors into the limb buds (in species that have them) which extend from the body wall (Chevallier et al 1977). Eventually, 30 (zebrafish), 65 (mouse), or as many as 500 (certain snakes) somites will form (Maroto et al 2012).

Cooke and Zeeman presented a mathematical model for somitogenesis in 1976, before any of the involved molecular components had been identified (Cooke and Zeeman 1976). It drew on two physical processes, intracellular biochemical oscillations, and a traveling morphogenesis-permissive signal sweeping across the length of the embryo. In this "clock-and-wavefront" mechanism, the unsegmented tissue at any position along the axis was postulated to be in a "phase-linked" (i.e., synchronized) state. When a cohort of cells that were at a certain phase of the cycle experienced the wavefront signal, they would coalesce into a somite. A spatial gradient of phase values, the slow unidirectional passage of the wave, or both, could ensure that somite pairs formed in the observed anteroposterior order.

While both the oscillatory and spatially progressive aspects of the Cooke-Zeeman model were speculative, periodic cell regulatory processes, most prominently the cell cycle, were long known. Sustained oscillations in energy metabolites had been observed a few years previously in yeast cells (Ghosh et al 1971), and their biochemical mechanism established in cell-free reactors (Pye and Chance 1966). The connection between such intracellular oscillations and the hypothesized synchronous oscillations that the somitogenesis model posited was obscure, but the authors were drawing on new theoretical initiatives on the dynamics of periodic phenomena in biological systems taking place during the same period (Winfree 1970; Goldbeter and Lefever 1972).

In the late 1990s Pourquié and coworkers appeared to establish the validity of the clock-and-wavefront model by identifying molecular components that played the predicted roles. In chicken embryos the product of the Hes1 gene is expressed in the paraxial mesoderm in cyclic waves whose period corresponded to the formation time of one somite (Palmeirim et al 1997; Pourquié 2003). This protein and its homolog Her7, another member of the Hes/Her family of transcriptional coregulators were found to be similarly expressed in zebrafish and mouse, and dynamical bases of the intracellular oscillation were inferred (Lewis 2003; Monk 2003). As noted above, however, self-organizing biochemical processes in individual cells do not constitute developmental self-organization, unless they come to act on the multicellular level. Lewis and Kageyama and their respective coworkers independently demonstrated that the Hes/Her oscillations become entrained so that, as predicted by Cooke and Zeeman and suggested by the experiments of Palmeirim et al, the PSM at each tissue domain about to undergo somite formation is a synchronized multicellular oscillator (Giudicelli et al 2007; Özbudak and Lewis 2008; Masamizu et al 2006). This collective behavior is uncontroversially the result of self-organization, but the way synchrony is brought about appears to differ in different species and remains in dispute even in well-studied organisms.

There are two possibilities. The first is that the cells are individually clock-like: they can keep time (i.e., undergo cyclic changes in Hes/Her concentration) even if isolated from one another, that is, autonomously, and can come into phase synchrony by interacting with each other in weak, non-specific ways. This so-called "Kuramoto effect" is a mathematically confirmed consequence of the physics of oscillations, and is observed, for example, in the behavior of fireflies in the wild and initially asynchronous arrays of metronomes that have been lined up on a flexible board (Strogatz 2003).

The second possibility is that the cells, though potentially oscillatory, are excitable entities that mainly enter this dynamical mode by interacting with each other. In this case, the establishment of clock-like behavior at the individual cell-level coincides with synchronization of the collective (Baibolatov et al 2009). It is this second phenomenon which is commonly referred to as dynamical self-organization, because it mediates an emergent synchronized state. But the Kuramoto mechanism, which brings coherence to a set of phase incoherent oscillations is also a form of physical self-organization by the definition used in this chapter. Recent experiments suggest that synchronization of PSM cells in some somite-forming species is due to the non-Kuramoto emergent type of self-organization (Hubaud et al 2017).

In intact embryos, somite boundaries form when cells which have left the posterior growth zone move sufficiently far from a source of FGF8 (and in some species additional, or different factors) in the posterior tip of the embryo (Dubrulle et al 2001; Mallo 2016). But when PSM is dissociated and placed in culture, somite-like structures form whose size and shape is controlled solely by local cell-cell interactions, involving neither a clock nor a wavefront (Dias et al 2014). This indicates that important aspects of somitogenesis result from inherent mechanical instabilities of the PSM.

Electron microscopic studies by Meier from the late 1970s suggested that, at least in chicken embryos, the relevant tissue is already arranged into somite-like cell clusters (termed "somitomeres") before the cells would have moved outside any posterior zone of morphogenetic inhibition (i.e., the wavefront of the Cooke-Zeman model) (Meier 1984, 1979). Further, experimental application of mechanical strain to living embryos was shown to induce supernumerary well-formed somites between existing ones, by a mechanism involving somite splitting and reorganization (Nelemans et al 2020). This provided evidence that mechanical organizational effects are operative in situ, even under normal stress conditions. Finally, newer electron microscopic observations revealed a posteriorly progressing front of adhesive changes in the presomitic mesoderm, accompanied by signs of tissue segmentation, developmentally much earlier than overt intersomite boundary formation (consistent with Meier's findings). The investigators have advanced a model of mechanical instability of the PSM incorporating this work, challenging the fundamentality of the clock-and-wavefront mechanism (Adhyapok et al 2021).

As with gastrulation, developmental processes in the embryo are much more complex than in isolated tissues, and it seems plausible that the clock and wavefront (with somewhat different molecular-genetic bases in different vertebrate species) evolved to reinforce a more primitive segmentation scheme that evolved in an ancestral form (Stern and Piatkowska 2015; Mallo 2016). (Strikingly, short germband insects like beetles and grasshoppers, though not directly ancestral to vertebrates, also employ clocks and wavefront-like activity gradients, using some of the same genes as the latter in their segmentation processes (Clark et al 2019).) Nonetheless, the cell aggregate-forming process that occurs in vitro, and which is essential to somite formation in the intact embryo, appears to be an authentic self-organizing process (Dias et al 2014).

Oscillator synchrony, by whichever means it is brought about (e.g., via the Kuramoto or emergent modes) also appears to be a generic self-organizing process. Its effect is to coordinate cell states across multicellular tissues in a variety of developing systems (see below). The generation of such "morphogenetic fields" (Levin 2012, Gilbert et al 1996) enable the reliable and coherent operation across broad tissue domains of system-specific spatiotemporal signals.

*Limb skeletogenesis*

The jawed vertebrates (gnathostomes) typically have two sets of paired appendages, either fins or limbs. These structures are characterized by species-specific arrangements of bony or cartilaginous endoskeletal elements that arise from focal condensations of "mesenchyme," i.e., loose embryonic connective tissue, similar to those observed in early somite formation. The mesenchymal cells are initially dispersed in a dilute extracellular matrix (ECM), but this material is lost or greatly reduced at the sites of condensation, and the cells come to directly adhere to one another. Mesenchymal condensation is a transient effect. The skeletal primordia progress to cartilage tissue in cartilaginous fish such as sharks where it forms the definitive skeleton, as well as in lobe-finned fish, such as coelacanths, and tetrapods, where it is mostly replaced by bone later in development. Replacement bone (e.g., that formed with the intermediary of cartilage) also occurs to a limited extent in ray-finned fish, but the bony endoskeleton in these species mostly develops by direct differentiation of mesenchymal condensations into bone in the so-called dermal rays of the fins (Clack 2012; Hinchliffe and Johnson 1980).

There were indications as early as the 1960s that self-organizing processes contribute to vertebrae limb skeletogenesis. Zwilling performed a series of experiments in which he disrupted the mesoblast from isolated chicken limb buds and packed the fragments or reaggregated cells into a jacket of ectoderm (i.e., embryonic skin) from a different limb bud. When the composite was grown as a graft, recognizable limb structures, consisting of nodules or small rods of cartilage, developed (Zwilling 1964). These experiments were repeated by Fallon and coworkers, with the addition that expression of Hox genes was monitored in the "recombinant" limbs (Ros et al 1994). According to the theory of "positional information," a developmental program-type model that was popular in limb research at the time, the distribution of various Hox gene products in the normal limb were what specified the locations of the skeletal elements (Tickle 1994). But the investigators found that discrete, parallel rod-like elements reminiscent of digits could appear without the reestablishment of normal Hox protein gradients. This provided further evidence that the Zwilling result was indeed due to some form of self-organization.

In tetrapod limbs and lobe-finned fish fins, the skeleton is arranged so that there is generally one element attached to the body, with increasingly distal rows containing more of them (see Fig. X.1, bottom panel, for the mouse case). In the human arm, for example, the proximal humerus is followed by the radius and ulna, two tiers of wrist bones, and then five digits. In birds and mammals these elements are generated proximodistally during development, so that rows containing successively larger numbers of elements (initially parallel rods of cartilage) emerge sequentially over time (Hinchliffe and Johnson 1980).

There are generic physical mechanisms of self-organization that permit well-spaced spots or stripes of chemical concentration or states of mechanical compression to form, with different numbers appearing as a result of fine-tuning of system parameters or changes in domain shape (Alber et al 2005). The Turing mechanism is one of these, and was proposed to underlie limb skeletal patterning beginning in the 1970s (Newman and Frisch 1979; Hentschel et al 2004). We can ask if any of these generic mechanisms underlie the patterning of the mesenchymal tissue of the limb mesoblast. The proximodistal generation of the different tiers of the skeleton does appear draw on such physical processes. But since the spatiotemporal ordering depends on regulated tissue elongation and the regulated decreasing suppression of pattern formation in the distal regions of the limb bud by proteins (mainly fibroblast growth factors; FGFs) secreted by the apical ectoderm, there are aspects of the process that while mechanistically comprehensible, are not themselves physically self-organizing (Glimm et al 2020, Zhang et al 2013).

Like somitogenesis, fin and limb skeletogenesis depends on patterns of mesenchymal condensation in which cells form focal aggregates that come to be distributed in a regular fashion across the developing tissue. There is, in fact, evidence for one or more Turing-type reaction-diffusion processes occurring in fin and limb development and in avian and mammalian limb bud cells in vitro (Christley et al 2007, Kiskowski et al 2004; Raspopovic et al 2014); Sheth et al 2012). But simulations show that fairly regular spot and stripe patterns can also result from local cell-cell and cell-ECM interactions in the absence of diffusible activators and inhibitors (Zeng et al 2003).. Unlike Turing processes, which require such factors, the simpler cell-cell and cell-ECM-based mechanism of Zeng and coworkers has no intrinsic wavelengths. Its capacity to generate regular patterns is conditional, dependent on the density of the cells and the strength of their local interactions (Zeng et al 2003). As in the case of somite formation, superimposed dynamical processes, some with self-organizing properties, which may differ in divergent lineages (e.g., cartilaginous fishes, birds, mammals) could serve to regularize the patterns formed by the simpler, likely more ancestral, mechanism.

In attempting to determine whether there is a self-organizing condensation-generating process common to all fin and limb development, it is helpful to compare experimental studies that have traced the development of condensations in vivo and in vitro. Although fibronectin, an ECM molecule that is abundant in avian and mammalian limb precartilage condensation is important in the consolidation of these foci as they mature (Frenz 1989, Downie and Newman 1994), and it is part of an elaborate reaction-diffusion system involving various growth factors and their receptors (Hentschel et al 2004), it is not the developmentally earliest molecule to mark these structures. A receptor for bone morphogenetic protein (BMP), a diffusible morphogen produced by the mesenchymal cells is present earlier when cell "compactions" or "protocondensations" appear in anticipation of the condensations (Barna and Niswander 2007). There is also evidence for a Turing-type network incorporating BMP in conjunction with another morphogen, Wnt, and Sox9, a master transcription factor for the differentiation of cartilage in the formation of digits in the developing mouse limb (Raspopovic et al 2014). However, this "BSW" network falls short as a candidate for the ancestral self-organizing system for fin and limb skeletogenesis.

In the first place, protocondensations form in the absence of Sox9, i.e., when it is knocked down genetically (Barna and Niswander 2007). This makes sense for the dermal endoskeleton of ray-finned fishes, where the bones form without a cartilage intermediate. While recent evidence suggests the dermal rays are organized in part by the same pattern-forming system as tetrapod limb endoskeletons (Gehrke et al 2015), here the master transcription factor of bone, Runx2, is involved rather than Sox9.

Second, there is no evidence that skeletal elements other than the digits are patterned by the BSW mechanism. A primordial self-organizing process that has been recruited across all fins and limb would be expected to function not only in the digital plate (which is a novelty of tetrapods) and its counterpart in the distal region of the developing fins of cartilaginous fish (Onimaru et al 2016), but at all proximo-distal levels of the limb, particularly those with deeper evolutionary roots than the digits (Stewart et al 2017).

Third, none of the components of the BSW network are cell-cell adhesive or ECM molecules. Factors that mediate changes in the arrangement of cells are prerequisites for the kind of ancestral condensation self-organizing system postulated above.

The one experimentally confirmed molecular mechanism that promotes condensations in limb mesenchyme consists of galectin-1 (Gal-1) and galectin-8 (Gal-8), two members of the galectin family of sugar-binding proteins (Bhat et al 2011). Galectins are matricellular proteins (i.e., ECM components that are rapidly turned over and play both adhesive and signaling roles; Gabius (2009)). In the avian limb bud tissues where the role of galectins has been studied, Gal-1 localizes to prospective sites of condensation where it promotes aggregation of limb mesenchymal cells, while Gal-8 blocks the ability of Gal-1 to perform this function. Though antagonistic at the protein level, the two galectins mutually induce each other's expression at the gene regulatory level. This combination of antagonistic and reinforcing interactions, along with the effects that each galectin has on the production (or mobilization at the cell surface) of its specific and shared ligands, constitutes a multiscale network capable of forming regular patterns (Glimm et al 2014).

Although the galectin-based mechanism can be described as a Turing-type system, it depends on local cell-adhesive cell-ECM interactions similarly to the mechanism proposed by Zeng et al (2003) rather than on a balance of diffusible morphogens and their effects, as in classical Turing mechanisms. Because Gal-8 interferes with cell-cell adhesion mediated by Gal-1, it is superficially analogous to the inhibitory morphogen in a reaction-diffusion mechanism. But it acts entirely locally. In a model based on its experimentally determined action, changing its expression rate and binding affinity to a postulated shared receptor with Gal-1 changes the pattern wavelength, but altering its diffusion rate does not (Bhat et al 2016; Glimm et al 2014). Further, the simulated galectin mechanism only forms patterns if the cells it acts on move up gradients of Gal-1, another feature not part of Turing mechanisms. This has led it to be called a "reaction-diffusion-adhesion" mechanism (Glimm et al 2014).

Cell movement in response to an adhesive gradient is a generic effect since it does not require the cells to be anything other than randomly motile. The galectin-based protocondensation-forming mechanism can thus be considered an example of tissue self-organization.

Although little is yet known about the role, if any, of the galectin network in non-avian species, homologs of Gal-1 and Gal-8 are present in all tetrapods and ray-finned and cartilaginous fishes for which genomic data are available. The sequence and conformation of Gal-8s in these groups provide a basis (insofar as basic computational models permit) for understanding the different numbers of limb skeletal elements in these species. Moreover, the acquisition of putative cis-acting regulatory sequences by the Gal-8 genes of tetrapods provides an explanation (by simulations) of the proximodistal increase in element number in this group (Bhat et al 2016). Thus, the two-galectin network, which may have recruited the lineage-determining transcription factors of cartilage (Sox9) or bone (Runx2) during the evolution of various fishes and tetrapods, is the best candidate for a primordial self-organizing basis of skeletogenesis (Newman et al 2018).

A second self-organizing process active during skeletogenesis resembles the synchronization of cellular oscillations employed in somitogenesis. As in the PSM, Hes1 oscillates in limb bud mesenchyme, and when these oscillations were suppressed in vitro, condensation patterning was perturbed (Bhat et al 2019). Depending on the time in culture when this was done there was either an increase in the number of protocondensations with less regular spacing between them, or a decrease and fusion of neighboring condensations. Suppression of Hes1 oscillations in the limb buds in intact embryos similarly led to pattern irregularities, including fused and misshapen skeletal elements (Bhat et al 2019).

A functional connection between Hes1 and the core two-galectin network motivated computational studies of the effects of perturbing variables representing the concentration of each protein on the dynamics of the others. When Hes1 was made not to oscillate, the simulations showed that the galectin system formed regularly spaced condensations with sharp boundaries, and when Hes1 oscillated in a synchronous fashion the result was the same.

However, if Hes1 was allowed to oscillate asynchronously, or its concentration was assigned randomly across the field of cells, the precision of the pattern was degraded.

The implication is that synchronization of oscillators, rather than being a pattern-forming mechanism itself, is a dynamical means to place all the cells in a developing field on the "same page." That is, if the levels of the multifunctional transcription coregulator Hes1 are uniform across the tissue, patterning signals such as nonuniform concentrations of galectins (in limb skeletogenesis) or the release from posterior suppression of aggregation (as in somitogenesis) can act in a spatiotemporally coherent fashion over distances greater than the range of the diffusible signals usually thought to coordinate development (Bhat et al 2019).

## X.4 Conclusions

Animal embryos and developing organs, being parcels of excitable soft matter, will inevitably exhibit physical self-organizing effects. Does that mean that embryos, or embryo- and organ-like structures derived from stem cells, are physically self-organizing? The studies described here suggest that they are not, and the most suitable description for what they do in cultures (as in their normal gestational environments) is just "development." As the examples have indicated, physical processes of self-organization are clearly employed in the embryos and organ primordia of extant animals: envelopment of the yolk by the blastoderm in the zebrafish embryo by tissue surface tension according to the differential interfacial tension mechanism, PSM condensation and limb/fin mesenchymal protocondensation compaction by mechanical instabilities. The multicellular synchronization of oscillations of transcriptional regulators is another self-organizing effect employed in embryogenesis and evidenced in vitro. But the constitution of the interface between the reorganizing germ layers in the fish embryo. the organization of PSM condensations into tandem somites, and the limb/fin condensations into patterned endoskeletons occur by agent-type and growth-dependent processes that are "biological" in ways not readily explicable by physical mechanisms.

Like the gastrulation example, the recruitment of simple nodule-generating self-organizing effects into patterned blocks of tissue along the body axis, or cartilaginous and bony fin or limb endoskeletons, required departures from their presumed ancestral states. Somitogenesis involved integration of tissue condensation into the elongating body in a way that suppresses it until it moves beyond the influence of an inhibitor emanating from the embryo's posterior tip. In all vertebrate species, the condensation process is under the control of Hes/Her family transcriptional regulators, which oscillate in concentration, so the resulting blocks of tissue are tandemly repetitive. But the posterior inhibitory factor is taxon-specific and varies in molecular mechanisms and timing, indicating phylogenetic divergence (Stern and Piatkowska 2015).

In the case of fins and limbs, the presumed ancestral protocondensation-initiating molecule Gal-1 may have generated randomly distributed supportive nodules in the appendages of ancient gnathostomes. But it then acquired embellishments that made the nodules more regular (coregulation with Gal-8), stronger (linkage to a reaction-diffusion network for fibronectin production), more solid (linkage to reaction-diffusion control of Sox9 or Runx2 for cartilage and bone, respectively) and finally, with acquisition by Gal-8 of cis-acting gene regulatory modules, position-dependent differentiation capabilities. This "evolving complex of self-organizing systems" generated, in a stepwise but divergent progression, the fins of cartilaginous and ray-finned fishes and the limbs of tetrapods (Newman et al 2018).

The described scenario posits major roles for physical processes of self-organization, not in the development of present-day embryos and organs, but in the origination of the forms and patterns of these forms. Since we cannot experiment on the ancestral forms to determine how they arose, the best evidence for this hypothesis is the "generic" look of some morphological features of the modern forms, that is, the appearance of having been formed by physical processes that are well understood via nonliving counterparts (Newman and Comper 1990). The configuration of embryonic germ layers, the segments of the vertebrate embryo, and the repeated endoskeletal elements of gnathostome appendages are examples of this. But as we have seen, whatever the means by which they first arose, evolution can also draw established morphological traits away from their original genetic and even physical foundations by the processes of "developmental system drift" (True and Haag 2001) and "homomorphy" (Newman 2019a).

The 18$^{th}$ century philosopher Immanuel Kant coined the term "self-organization" to describe the enigmatic properties of organisms. He described them as "natural purposes," entities that not only maintain coherent identities, but produce the ingredients and embody the means to perpetuate themselves (Kant 1790; trans. 1966). Kant wrote in the context of Newtonian physics, unaware of the thermodynamic and dynamical phenomena that later scientists would term self-organizational. Since many of these self-organizing processes operate in living systems, Kant's usage was enormously prescient.

But Kant was also only vaguely aware of organic evolution, and it is this, as we have seen, that transforms organisms into something different from the self-organizing systems of physics. Indeed, the complex, overdetermined entities that have resulted from evolutionary processes occurring over the more than 600 million years since the origin of the animals are, despite Kant's lack of appreciation of this dimension, aptly described by his term "natural purposes" (Moss and Newman 2015).

Finally, to return to the recent flush of allusions to self-organization, mentioned at the beginning of this chapter, in reference to formation of gastruloids and organoids produced by stem cells in culture (Beccari et al 2018; Etoc et al 2016; Rosado-Olivieri and Brivanlou 2021; Shahbazi et al 2019), it is difficult not to conclude that it is problematic, both scientifically and regarding social uses of the associated technologies. One recent paper, for example, describes what occurs when stem cells generate embryo-like structures in culture as akin to coalescence in a supersaturated vapor, phase separation of immiscible liquids, Turing processes, and so forth (Shahbazi et al 2019). Public lectures by principals of the field contain confidently asserted, but similarly oversimplified or misleading treatments (Brivanlou 2016). These analogies (understandable in the past, but less so now that the physics of some developmental systems have been investigated in depth, with documented complicating effects) give the false impression that there has been more progress in understanding embryonic development than there truly has been. Based in part on such "hand-waving" explanations, investigators are drawing inferences about human development and its genetic and environmental vulnerabilities from in vitro models, seeking to mobilize medically targeted funding to this end (Clark et al 2021). Notwithstanding the hazards of genetically modifying systems whose principles of organization are poorly understood (Newman 2017), some are forthright in their intention to eventually perform such alterations in human embryos based on data and purported conceptual advances acquired from stem cell development in vitro (Turocy et al 2021).

As we have seen, however, development of the embryo its in vivo setting does not generally occur by the action of physically straightforward processes, and development of reconstituted embryo cells in vitro does not always occur in the same way it does in vivo. It would be unfortunate if we find ourselves having emerged from a period of misconceived genetic program metaphors only to land in a brave new world captivated by equally misguided ones about self-organization.

**Figure legend**

**Fig 1** The three embryonic systems discussed in this chapter. Top panel, drawings of successive stages (left to right) of gastrulation in a zebrafish embryo, beginning from the formation of the interface between the germ layers at 4 h postfertilization to the envelopment of the yolk cell layer by the blastoderm at 10 h postfertilization. Reprinted from Bruce and Heisenberg (2020) Mechanisms of zebrafish epiboly: A current view. *Curr Top Dev Biol* 136: 319-341, used by permission of Elsevier. Middle panel, photographs of chicken embryos during the period of somitogenesis (26-53 h; 8-22 somites). Adapted from Hamburger and Hamilton (1951). Bottom panel, drawing of cross-sections of upper and lower embryonic mouse limbs between 9.5 and 13.5 d of development. Gray shading represents precartilage mesenchymal condensations, and black shading cartilage. Adapted from Taher et al (2011).

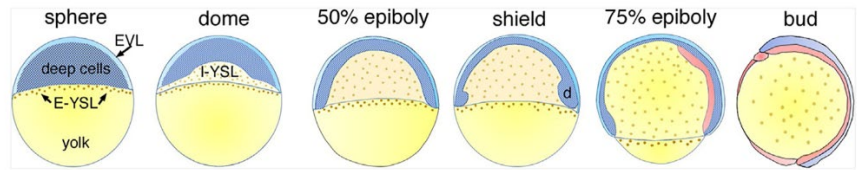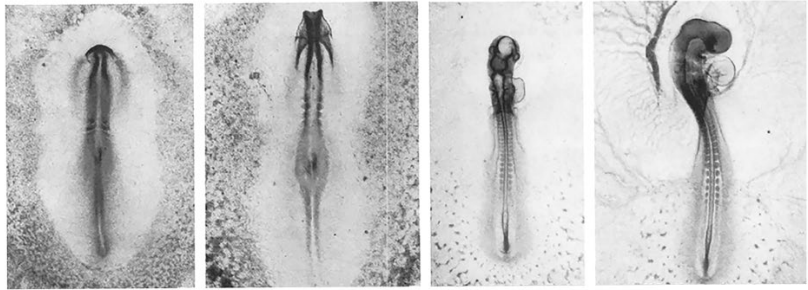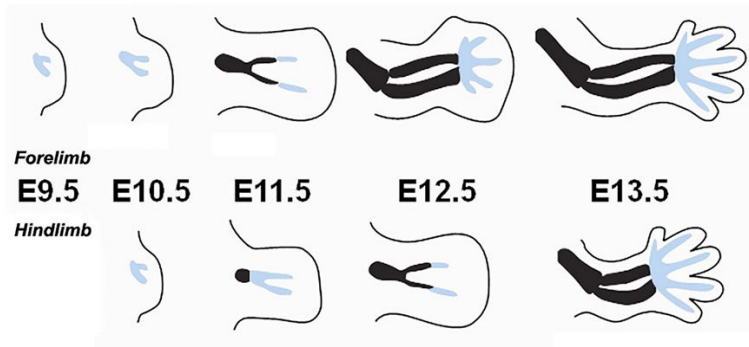

**Fig. 1**

36